\documentclass[runningheads]{svmult}
\newcommand{\bit}{\begin{itemize}}
\newcommand{\eit}{\end{itemize}}
\newcommand{\bea}{\begin{eqnarray}}
\newcommand{\eea}{\end{eqnarray}}
\newcommand{\be}{\begin{equation}}
\newcommand{\ee}{\end{equation}}
\newcommand{\f}{\frac}
\newcommand{\e}{{\rm e}}
\newcommand{\kav}{\left<k\right>}

\usepackage{graphicx}    
\usepackage{multicol}    

\makeindex             

\begin{document}
\newcommand{\mytitle}{Equilibrium statistical mechanics of network structures}
\newcommand{\myshorttitle}{Equilibrium statistical mechanics of network structures}
\title*{\mytitle}
\toctitle{\mytitle}
\titlerunning{\myshorttitle}
\authorrunning{I. Farkas\and
I. Der{\'e}nyi\and 
G. Palla\and 
T. Vicsek
}

\author{I. Farkas$^{\dagger}$\and
I. Der{\'e}nyi$^{\ddagger}$\and 
G. Palla$^{\dagger}$\and 
T. Vicsek$^{\dagger,\ddagger}$
}
\institute{
$^{\dagger}$Biological Physics Research Group of HAS and $^{\ddagger}$Department\ of Biological Physics, E\"otv\"os University,
 P\'azm\'any P.\ stny.\ 1A, H-1117 Budapest, Hungary
}

%
\maketitle

\begin{abstract}
In this article we give an in depth overview of the
recent advances in the field of equilibrium networks. 
After outlining this topic,
we provide a novel way of defining equilibrium graph (network) ensembles.
We illustrate this concept on the classical random graph model and then 
survey a large variety of recently studied network models.
Next, we analyze the
structural properties of the graphs in these ensembles 
in terms of both local and global characteristics, such as degrees,
degree-degree correlations, component sizes, and spectral properties.
We conclude with topological phase transitions 
and show examples for both continuous and discontinuous transitions.
\end{abstract}

\section{Introduction}
\label{sec:intro}

A very human way of interpreting our complex world 
is to try to identify subunits in it
and to map the interactions between these parts.
In many systems, it is possible to define subunits 
in such a way that the 
network of their interactions
provides a simple but still informative representation of the system. 
The field of discrete mathematics dealing with networks 
is {\it graph theory}.

Research in graph theory
was started by Leonhard Euler \cite{euler}.
In the 1950s another major step was taken by
Erd\H os and R\'enyi: they introduced the notion of 
classical random graphs 
\cite{erdos-renyi,gilbert,bollobas-book}.
By the late 1990s more and more actual maps of large networks 
had become available and modeling efforts were directed
towards the description of the newly recognized properties of 
these systems
\cite{watts-strogatz,sf,albert-revmod,dm-book,bornholdt-book}.
A network is constructed from many similar subunits (vertices)
connected by interactions (edges),
similarly to the systems studied in statistical physics.
Because of this analogy,
the methods by which some of the central problems of statistical
physics are effectively handled, 
can be transferred to networks, e.g.,
to graph optimization and topological phase transitions.

In this article we will discuss the construction
of network ensembles that
fit into the concept of equilibrium as 
it is used in statistical physics,
with a focus on structural transitions
\cite{berg-laessig,manna-hamiltonian,dms-principles,derenyi,palla}.
Note that even though structural transitions in growing networks 
are non-equilibrium phenomena
\cite{szabo-structural,bianconi-condensation,krapivsky-gelation},
some of the main features of the structures constructed by growth can
be reproduced by non-growing models
(see, e.g., Refs. 
\cite{manna-hamiltonian,dms-principles,bollobas-diameter}).
Similarly, non-growing graphs are not 
necessarily equilibrium systems (see, e.g., Ref. \cite{dm-book}).
Closely related real-world phenomena and mathematical models are
the configurational transitions of branched polymers
\cite{bialas-transition},
structural transitions of business networks during changes of the
``business'' climate
\cite{stark-vedres,onnela-taxonomy},
the transitions of collaboration networks \cite{newman-collab},
networks defined by the potential 
energy landscapes of small clusters of atoms \cite{doye},
and potentials on tree graphs as introduced 
by Tusn{\'a}dy \cite{tusnady-trees}.

In the present review we intend to go beyond those 
that have been published previously
\cite{dm-book,dms-principles,burda-statmech}, 
both concerning the
scope and the depth of the analysis.

We will {\it focus on the structure of networks}, 
represented by graphs,
and will not consider any dynamics {\it on} them.
Thus, several widely studied models 
are beyond the scope of the present review:
models using, e.g., spins on the vertices 
\cite{dms-principles,gdm-critical,dorog-Ising,bianconi-Ising,barrat-SW,kim-XY,dorog-Potts},
disease spreading 
\cite{newman-disease,zanette-immunization,zekri-disease}
agent-based models on networks
\cite{toro-competition},
or weighted edges and traffic on a network
\cite{yook-modeling,pastor-dynamical,guimera-optimal,braunstein-optimal}.

{\it Definiton:}
Natural networks mostly arise from non-equilibrium processes,
thus, the notion of equilibrium in the case of networks 
is essentially an abstraction 
(similarly to any system assumed to be in
perfect equilibrium).
We {\it define equilibrium network ensembles} as stationary
ensembles of graphs 
generated by restructuring processes obeying {\it detailed balance}
and ergodicity.
During such a restructuring process, edges of the graph are
removed and/or inserted.

This definition raises a few issues to be discussed. 
First of all, the characteristic timescale of rewiring one particular link
varies from system to system. For example, the network 
of biochemical pathways \cite{kohnmap}
available to a cell can undergo structural changes 
within years to millions of years, in contrast to  business interactions
\cite{stark-vedres,newman-collab}, which 
are restructured over time scales of days to years, 
while the characteristic times of technological networks may be even
shorter. 
With a finite number of measurements 
during the available time window
it is often difficult to decide whether
a graph that has not been observed 
has a low probability or it is not allowed at all.
Hence, the set of allowed graphs is often unclear.
A simple way to by-pass this problem is to enable
all graphs and tune further parameters of the model
to reproduce the statistics of the observed typical ones.

Once the set of allowed graphs has been fixed, the 
next step in the statistical physics treatment of a network ensemble
is to fix some of the thermodynamic variables, e.g.,
for the canonical ensemble one should fix the
temperature and all extensive variables except for the 
entropy\footnote{In the mathematics literature, 
the entropy of graphs has been analyzed in detail 
\cite{simonyi-entropy,csiszar,bauer-entropy}.}.
At this point, an {\it energy function} would be useful. 
Unfortunately, unlike in many
physical systems, the energy of a graph 
cannot be derived from first principles. 
A possible approach for deriving an energy function is 
reverse engineering: one tries to reproduce the 
observed properties of real networks 
with a suitable choice of the energy in the model.
Another possibility can be to explore the effects of
a wide range of energy functions on the structures of networks.
Alternatively, 
to suppress deviations from a prescribed target property,
one can also introduce a cost function (energy).
Having defined the energy, one can proceed towards a
detailed analysis of the equilibrium 
system using the standard methods of statistical physics.

Often a complete analogy with statistical physics is
unnecessary, and shortcuts can be made to simplify the above
procedure. It is very common to define graph 
ensembles by assigning a statistical weight to each allowed graph,
or to supply a set of master equations describing the dynamics of
the system, and to find the stable fixed point of these equations.
Of course, skipping, e.g., the definition of the energy will
leave the temperature of the system undefined. 

This article is organized as follows. In Section \ref{sec:prelim}
we introduce the most important notions.
Section \ref{sec:ensembles} 
will concentrate on currently used graph models and 
the construction of equilibrium graph ensembles.
Section \ref{sec:features} will discuss some of 
the specific properties of these sets of graphs.
In Section \ref{sec:tpt} examples will be given 
for topological phase transitions of graphs
and Section \ref{sec:sum} contains a short summary.

\section{Preliminaries}
\label{sec:prelim}

Except where stated otherwise, 
we will consider {\it undirected simple graphs}, i.e., 
non-degenerate graphs where
any two vertices are connected by zero or one undirected edge, 
and no vertex is allowed to be connected to 
itself\footnote{In a degenerate (or pseudo) graph multiple 
connections between two vertices
and edges connecting a vertex to itself are allowed.
Some additional extensions are
to assign, e.g., weights and/or fitnesses to
the edges and vertices.}.
The number of edges connected to the $i$th 
vertex is called the degree, $k_i$, of that vertex. Two vertices
are called neighbors, if they are connected by an edge. 
The degree sequence of a graph is the ordered 
list of its degrees, and the degree distribution
gives the probability, $p_k$, for a randomly selected vertex
to have degree $k$.
The degree-degree correlation function, $p(k,k')$, gives the
probability that one randomly selected end point of a randomly chosen 
edge will have the degree $k$ and the other end point the degree $k'$.

The clustering coefficient of the $i$th vertex is the 
ratio between the number of edges, $n_i$, connecting
its $k_i$ neighbors and the number of 
all possible edges between these neighbors:
\bea
C_i = \f{n_i}{k_i(k_i-1)/2} \, .
\label{eq:clustering-coeff}
\eea
\noindent
The clustering coefficient of a graph
is $C_i$ averaged over all vertices.
The shortest distance, $d_{i,j}$,
is defined as the 
smallest number of edges that lead from vertex $i$ to $j$.
Finally, a set of vertices connected to each other by edges 
and isolated from the rest of the graph
is called a {\it component} of the graph.

The two basic constituents of a simple graph 
are its vertices and edges, therefore it is essential 
whether a vertex (or edge) is distinguishable from the others.
In this article, we will consider {\it labeled graphs}, i.e., 
in which both vertices and edges are distinguishable.
A graph with distinguishable vertices 
can be represented by its adjacency 
matrix, $\mathbf{A}$.
The element $A_{ij}$ denotes the number
of edges between vertices $i$ and $j$ if $i\not= j$,
and twice the number of edges if $i=j$ (unit loops).
For simple graphs, this matrix is symmetric,
its diagonal entries are $0$, 
and the off-diagonal entries are $0$ or $1$. 
Note, that the adjacency matrix is insensitive to whether 
the edges of the graph are distinguishable: 
swapping any two edges will result 
in the same $\mathbf{A}$.

Also, it is possible to define equivalence 
classes of labeled graphs using graph isomorphism:
two labeled graphs are equivalent, if there exists a permutation of the
vertices of the first graph transforming it into the second one.
As a consequence, each equivalence class of labeled graphs can be
represented by a single unlabeled graph 
(in which neither the edges nor the vertices are distinguishable). 
These equivalence classes will be referred to as
{\it topologies},
i.e., two graphs are assumed to have the same topology,
if they belong to the same equivalence class.
This definition is the graph theoretical equivalent of the definition of
topology for geometrical objects, where two objects have the same
topology, if they can be transformed into each other through
deformations without tearing and stitching.

The focus of this article is on graph restructuring processes.
Denoting the transition rates between graphs $a$ and $b$ by
$r_{a\to b}$, 
the time evolution of the probability of the graphs
in the ensemble can be written as a set of master equations:
\bea
\f{\partial P_a}{\partial t} = \
\sum_b ( P_b r_{b\to a} - P_a r_{a\to b} ) \, ,
\eea
where $P_a$ is the probability of graph $a$. 

If the dynamics defined in a system has a series of non-zero
transition rates between any two graphs (ergodicity), 
and there exists a stationary distribution, 
$P_a^{\mathrm{\,stat}}$ fulfilling the
conditions of detailed balance,
\bea
P_a^{\mathrm{\,stat}} r_{a\to b} = \
P_b^{\mathrm{\,stat}} r_{b\to a} \, ,
\label{eq:detailed-balance}
\eea
\noindent
then the system will always converge 
to this stationary distribution, 
which can thus be called 
{\it equilibrium distribution}.

In the reverse situation,
when the equilibrium distribution is given,
one can always create a dynamics that leads to this distribution.
Such a dynamics must fulfill the conditions of detailed balance 
and ergodicity. Since the detailed balance condition
(\ref{eq:detailed-balance}) fixes only the ratio of 
the rates of the forward and backward transitions between each
pair of graphs ($a$ and $b$), 
the most general form of the transition rates 
can be written as
\bea
r_{a\to b}=\nu_{ab} P_b \, ,
\label{eq:wij-pj}
\eea
\noindent
where all $\nu_{ab}=\nu_{ba}$ values are arbitrary factors
(assuming that they do not violate ergodicity).

\section{Graph ensembles}
\label{sec:ensembles}

Similarly to Dorogovtsev et. al
\cite{dm-book,dms-principles} 
and Burda et. al \cite{burda-ensemble}, 
we will discuss graph ensembles in this section.
According to statistical physics, 
for a rigorous analysis one needs to define 
the microcanonical, canonical, and grand canonical ensembles.
However, even if some of the necessary variables, 
(e.g., the energy) are not defined, 
it is still possible to define similar graph ensembles.

In equilibrium network ensembles, 
the edges (links) represent particles
and one graph corresponds to one state of the system.
In this article we will keep the number of vertices constant,
which is analogous to the constant volume constraint.

\subsection{Ensembles with energy}
\label{subsec:energy}

Energy is a key concept in {\it optimization problems}.
Even if it is not possible to derive an energy for graphs from
first principles, 
one can find analogies with well-established systems, 
and also phenomenological and heuristic arguments 
can lead to such energy functions
\cite{berg-laessig,manna-hamiltonian,dms-principles,derenyi,palla},
as described in the Introduction.

\subsubsection{Microcanonical ensemble.}
\label{subsubsec:energy-MC}

In statistical physics, the microcanonical ensemble
is defined by assigning identical weights to each state of a system
with a given energy, $E$, and a given number of particles;
all other states have zero weight.
Thus, the definition of a microcanonical ensemble is straightforward:
assign the {\it same} weight,
\bea
P^{\mathrm{\,MC}}&=&n^{-1} \, ,
\eea
\noindent
to each of the 
$n$ graphs that has $M$ edges and energy $E$,
and {\it zero} weight to all other graphs.

\subsubsection{Canonical ensemble.}
\label{subsubsec:energy-C}

The canonical ensemble is composed of graphs with a fixed number of
edges, and each graph $a$ has a weight
\bea
P^{\mathrm{\,C}}_a=\f{\e^{-E_a/T}}{Z^{\mathrm{\,C}}} \, ,
\label{eq:pc-x-beta}
\eea
where $T$ is the temperature,
$E_a$ is the energy of this graph, and 
\bea
Z^{\mathrm{\,C}}=\sum_b \e^{-E_b/T} 
\eea
denotes the partition function.
Network ensembles with a constant edge number 
and a cost function to minimize
the deviations from a prescribed feature
(e.g., a fixed total number of triangles),
belong to this category.

\subsubsection{Grand canonical ensemble.}
\label{subsubsec:energy-GC}

The grand canonical ensemble is characterized by a fixed temperature 
($T$)
and a fixed chemical potential ($\mu$).
The energy and the number of edges (particles) 
can vary in the system, and the probability of graph $a$ is
\bea
P^{\mathrm{\,GC}}_a=
\f{\e^{-(E_a-\mu M_a)/T}}{Z^{\mathrm{\,GC}}} \, ,
\label{eq:e-x-beta-mu}
\eea
where $E_a$ and $M_a$ denote the energy and edge number of graph $a$
respectively, and 
\bea
Z^{\mathrm{\,GC}}=\sum_b \e^{-(E_b-\mu M_b)/T}  
\eea
is the partition function.

\subsection{Ensembles without energy}
\label{subsec:without-energy}

\subsubsection{Microcanonical ensemble.}

Numerous network models are defined
through a static set of allowed graphs,
and no restructuring processes are involved.
Even if no energies and no probabilities are provided for these graphs, 
the microcanonical ensemble can still be defined
by assigning equal weight to 
each allowed graph \cite{dm-book,dorog-correq}.
This is equivalent to assigning the same 
energy to each allowed graph 
(and a different energy to all the others).

\subsubsection{Canonical ensemble.}

If a graph model provides probabilities, $\{P_a\}$,
for a set of graphs with an identical number of edges, 
then it can be considered as a canonical ensemble.
One can easily construct an energy function from the probabilities
using Eq. (\ref{eq:pc-x-beta}):
\bea
E_a = - T \log P_a + \log Z\, .
\label{eq:e-x-beta}
\eea
That is, the energy can be defined up to 
a factor, $T$, and an additive term, $\log Z$.

\subsubsection{The grand canonical ensemble.}

This ensemble is very similar to the canonical ensemble except that 
even the number of edges is allowed to vary. 
In this case an energy function can be constructed from
Eq. (\ref{eq:e-x-beta-mu}):
\bea
E_a = - T \log P_a + \mu M_a + \log Z\, ,
\eea
and a new, arbitrarily chosen parameter, $\mu$, appears.

\subsection{Basic examples}

\subsubsection{The classical random graph.}
\label{subsubsec:classical-random-graph}

We will discuss this classical example 
{\it to illustrate the concept} 
of equilibrium network ensembles.
The classical random graph model
is based on a fixed number ($N$) of vertices. 
The model has two variants.
The first one \cite{erdos-renyi} is the
${\mathcal{G}}(N,M)$ model:
$M$ edges are placed randomly and 
independently between the vertices of the graph.
The second variant \cite{gilbert} is the 
${\mathcal{G}}(N,p)$ model:
each pair of vertices 
in the graph is connected via an edge 
with a fixed probability, $p$.
In both variants
the degree distribution converges to a Poisson distribution
in the $N\to\infty$ limit:
\bea
p_k \to {\kav^k e^{-\kav}\over k!} \, ,
\label{eq:ERpk}
\eea
\noindent
where $\kav=2M/N$ in 
${\mathcal{G}}(N,M)$ and $\kav=pN$ in 
${\mathcal{G}}(N,p)$.
Viewing the edges as particles, the constant edge number variant of 
the classical random graph model
corresponds to the microcanonical ensemble,
since each particular configuration is generated with the same
probability.
In the constant edge probability variant,
only the expectation value of 
the number of ``particles'' is constant, 
and can be described by the grand canonical ensemble.

\begin{figure}[t!]
\centerline{\includegraphics[angle=0,width=0.9\columnwidth]{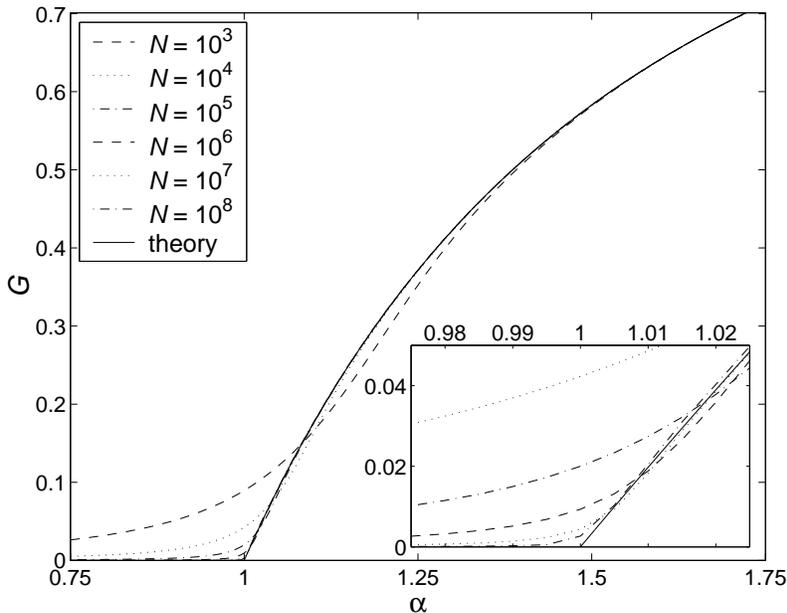}}
\caption{
Size of the largest component in a classical random graph as a
function of the average degree, $\alpha=\kav$, of a vertex. Note that
for $N>10^6$ the Monte Carlo data is almost indistinguishable from the
theoretical result in Eq. (\ref{eq:galpha}). 
Error bars are not shown,
because in all cases the error is smaller than the width of the lines.
The inset shows the transition in the vicinity of the percolation
threshold, $\alpha_{\mathrm{c}}=1$. 
Figure from Ref. \cite{dall-geometric}.
}
\label{fig:dall}
\end{figure}

At this point one should also mention the notion of the
{\it random graph process} \cite{erdos-renyi,bollobas-book},
a possible method for generating a classical random graph.
One starts with $N$ vertices,
and adds edges sequentially to the graph 
at independent random locations.
In the beginning, there will be many small components in the graph,
but after a certain number of inserted edges 
-- 
given by the critical edge probability, $p_{\mathrm{c}}$ 
-- 
a giant component\footnote{Note that 
the growth rate of this component is sublinear:
it grows as $\mathcal{O}(N^{2/3})$ \cite{ben-naim}.
} 
will appear.
This transition is analogous to 
percolation phase transitions. 
The fraction of nodes belonging to the largest component
in the $N\to\infty$ limit is \cite{bollobas-book}
\bea
G(\kav) = 1 - \f{1}{\kav} \sum_{n=1}^{\infty}
\f{n^{n-1}}{n!}(\kav e^{-\kav})^n \, .
\label{eq:galpha}
\eea
\noindent
This analytical result and actual numerical data
\cite{dall-geometric} showing the appearance of the giant component
are compared in Fig. \ref{fig:dall}.

\subsubsection{The small-world graph.}

Another well-known example for a graph ensemble
is the small-world model
introduced by Watts and Strogatz \cite{watts-strogatz}.
The construction of a small-world graph
starts from a one-dimensional
periodic array of $N$ vertices. 
Each vertex is first connected to its $k$ nearest
neighbors, where $k$ is an even positive number.
Then, each edge is moved with a fixed
probability, $r$, to a randomly selected new location. 
This construction leads to a canonical ensemble:
the number of edges is constant
and the probabilities of the individual graphs in the ensemble 
are different, because the number of rewired edges can vary.

\subsubsection{Ensembles with a fixed degree distribution.}
\label{subsec:fixed-degseq}

Many real-world graphs have a degree distribution
that decays slowly,
as a power law, as measured and described by 
Barab{\'a}si, Albert and Jeong \cite{sf,sf2}.
These graphs 
are often referred to as {\it scale-free}.
On the other hand, the classical random graph's
degree distribution has a quickly decaying ($1/k!$) tail
(see Eq. (\ref{eq:ERpk})).
The degree distributions of graphs have become
central to numerous analyses and various graph ensembles  
with fixed degree distributions have been developed
\cite{dm-book,dms-principles,burda-ensemble}.

\begin{figure}[t!]
\centerline{\includegraphics[angle=-90,width=0.9\columnwidth]{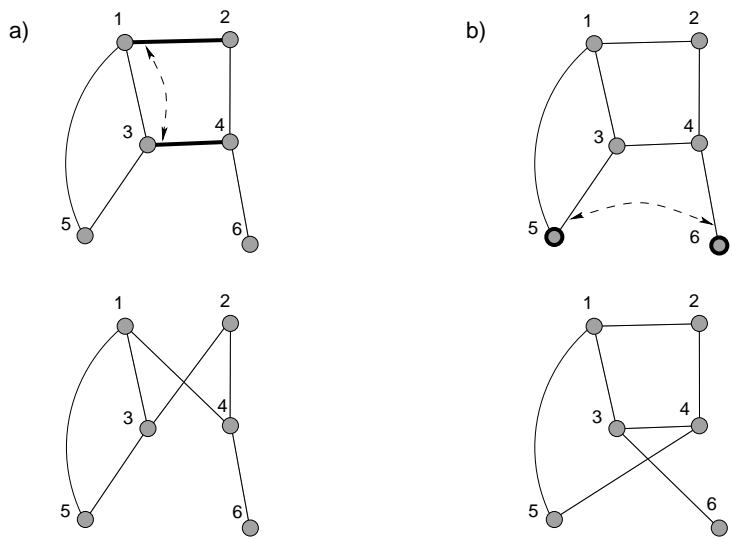}}
\caption{
Generating graph ensembles by randomization methods 
that leave the degree sequence of a graph unchanged.
(a) 
Link randomization. 
First, one selects two edges of the graph 
randomly. These are indicated by heavier lines: 
edges $1-2$ and $3-4$.
Then, one end point of each
edge is selected randomly 
-- the end points at vertices $1$ and $3$, respectively --
and the selected end points of these two edges are swapped.
(b) 
Vertex randomization. 
One starts with selecting two vertices at
random (vertices $5$ and $6$ in the example).
Next, one of the edges at each vertex is 
picked randomly and their end points
at the selected vertices are swapped.
}
\label{fig:vlrand}
\end{figure}

Given a network with the degree distribution $p_k$, 
there exist several re\-wiring algorithms 
that retain the degrees of all nodes at each rewiring step and 
generate an equilibrium ensemble of graphs.
Two examples are the
{\it link randomization}
\cite{maslov-specificity}
and the 
{\it vertex randomization}
\cite{xulvi-brunet-correlations}
methods. 
In both methods,
two edges are selected first, and then
one of the end points of each edge is picked and swapped,
ensuring that none of the degrees are changed. 
The two methods are explained in detail in Fig. \ref{fig:vlrand}.
The resulting canonical
ensembles will have the degree distribution $p_k$ in common,
but can have different equilibrium weights for the 
individual graphs. 
As pointed out by Xulvi-Brunet et. al \cite{xulvi-brunet-correlations},
upon link randomization
the degree-degree correlations are removed from a network,
but vertex randomization builds up positive degree-degree
correlations.

\subsection{Examples for graph energies}

\subsubsection{Energies based on vertex degrees.}

The most obvious units in a graph are the vertices themselves.
Therefore, it is
plausible to assign the energy to each vertex separately:
\bea
E = \sum_{i=1}^{N} f(k_i) \, .
\eea
\noindent
Note that if the number of edges is constant, then
the linear part of $f$ is irrelevant
(since its contribution is 
proportional to the number of edges in the graph), 
and simply renormalizes the chemical potential
in case of the grand canonical ensemble.
In the infinite temperature limit any 
$f$ will produce the classical random graph ensemble.
If $f$ decreases faster than linear, e.g., quadratically,
\bea
E=-\sum_{i=1}^{N} k_i^2 \, ,
\label{eq:ksqr-energy}
\eea
\noindent
then at low temperatures the typical graphs will 
have an uneven distribution of degrees among the vertices:
a small number of vertices with high degrees and 
a large number of vertices with low degrees. 
In a model of Berg et. al \cite{berg-laessig},
to avoid the occurence of isolated vertices 
and vertices with large degrees,
the following energy was proposed:
\bea
E=\sum_{i=1}^{N} \bigg[ -\f{k_i^2}{2}+\eta k_i^3\bigg] \, ,
\label{eq:berg-laessig-energy}
\eea
\noindent
and graphs containing vertices of zero degree were not allowed.

\subsubsection{Energies based on degrees of neighboring vertices.}

\begin{figure}[t!]
\centerline{\includegraphics[angle=-90,width=.8\columnwidth]{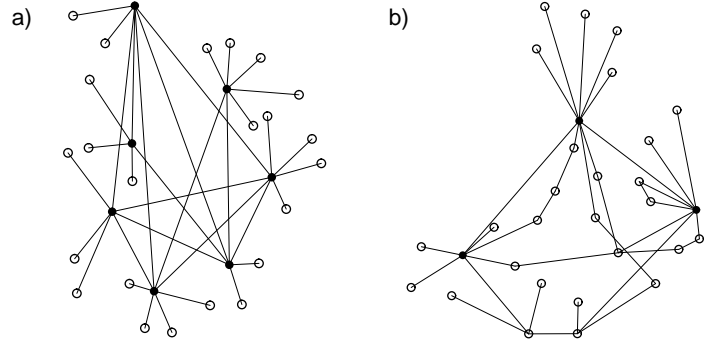}}
\caption{
Optimized networks generated by Berg and L{\"a}ssig 
\cite{berg-laessig} using 
(a) energies with local correlations, see
Eq. (\ref{eq:Berg-NNenergy}), and
(b) energies based on global properties, see
Eq. (\ref{eq:Berg-d-energy}).
In both cases, the temperature, $T$, was low.
Notice that in both graphs disassortativity is present
(see Sec. \ref{subsec:local-corr}):
vertices with high degrees 
(hubs, indicated by filled circles)
are preferentially connected to vertices with low degrees
(empty circles).
Figure from Ref. \cite{berg-laessig}.
\label{fig:BergLaessigFIG} 
}
\end{figure}

\begin{figure}[!t]
\centerline{\includegraphics[angle=-90,width=0.98\columnwidth]{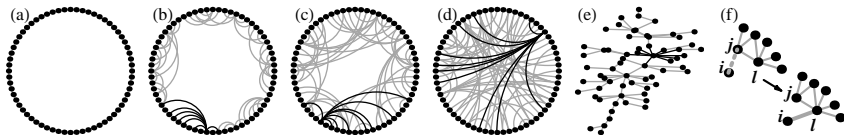}}
\caption{
(a)-(d)
Snapshots of the simulation used 
by Baiesi and Manna \cite{manna-hamiltonian}
to generate an ensemble of scale-free equilibrium
networks from a Hamiltonian dynamics, 
and (e)-(f) the Monte-Carlo 
rewiring method used during the simulation.
The initial network is constructed from 
$N$ links and $N$ vertices:
the vertices are connected as a ring. Later, 
$M-N$ ($M>N$) further edges are added to the network and 
the Monte-Carlo dynamics with the energy of Eq. (\ref{eq:E-manna})
is used to decide whether a randomly
selected edge, connecting vertices $i$ and $j_1$,
should be rewired to connect the previously unconnected pair of
vertices, $i$ and $j_2$.
Blue (dark) edges meet at the vertex with the highest degree.
Figure from Ref. \cite{manna-hamiltonian}.
\label{fig:Manna-Hamiltonian}
}
\end{figure}

Energies can also be assigned to edges,
\bea
E=\sum_{(i,j)} g(k_i,k_j) \, ,
\label{eq:Berg-NNenergy}
\eea
\noindent
where the summation goes over pairs 
of neighboring vertices (i.e., over the edges).
Energy functions of this type inherently lead to correlations
between vertices, as demonstrated by 
Berg and L{\"a}ssig \cite{berg-laessig} using
\bea
g(k_i,k_j)=\zeta\,\delta_{k_i,1}\delta_{k_j,1} \, ,
\eea
\noindent
see Fig. \ref{fig:BergLaessigFIG}.
Another example for this type of energy is
\bea
g(k_i,k_j) = \f{\min(k_i,k_j)}{\max(k_i,k_j)} - 1\, ,
\label{eq:E-manna}
\eea
which favors different degrees at the end points of an edge
\cite{manna-hamiltonian} (see Fig. \ref{fig:Manna-Hamiltonian} 
and Sec. \ref{subsec:local-corr}).

To account for correlations over longer distances,
a logical next step would be to 
add terms containing second neighbor interactions, e.g.,
\bea
E = -\f{1}{6}\, {\mathrm{Tr}}\,\mathbf{A}^3 \,  ,
\eea
\noindent
which counts the number of triangles in the 
graph with a negative sign.
If the number of edges can be written as $M=n(n-1)/2$ with an integer
$n$ ($n<N$), then  at low temperatures this energy 
leads to a complete (fully connected) subgraph on $n$
vertices, leaving the rest of the vertices ($N-n$) isolated.

\subsubsection{Energies based on global properties.}

The most apparent global properties of a network
are the sizes\footnote{Component 
  sizes are usually defined as the
  number of vertices in a component,
  however, in this article, because of the 
  edge--particle analogy,
  $s_i$ is the number of edges in the $i$th component.
}
of its components, and especially, 
the size of the largest component, $s_{\mathrm{max}}$.
A simple form of an energy containing 
component sizes is \cite{derenyi}
\bea
E=\sum_{i=1}^{n} f(s_i) \, ,
\eea
\noindent
where $n$ is the number of components in the graph and $s_i$ is the
size of the $i$th component.

The simplest form of the energy is 
proportional to the size of the largest component 
\bea
E=-s_{\mathrm{max}} \, .
\eea
In the ensemble defined by this energy 
as the temperature is lowered 
a phase transition occurs 
which is analogous to the density dependent transition
of the classical random graph 
(see Sec. \ref{subsubsec:classical-random-graph}).
This linear energy function was found to give a continuous transition
\cite{palla}, 
and the quadratic, 
$E=-s_{\mathrm{max}}^2$ or $E=-\sum_{i=1}^{N} s_i^2$
energies result in discontinuous transitions.

A possible goal of optimization 
can be to decrease the graph's diameter.
This can be realized with, e.g., the energy 
\cite{berg-laessig}
\bea
E=\sum_{i,j} d_{i,j} \, ,
\label{eq:Berg-d-energy}
\eea
\noindent
where the summation goes over all pairs of vertices.
See also Fig. \ref{fig:BergLaessigFIG} for a typical network generated
with this energy function.

\subsection{Mapping the graph onto a lattice gas}
\label{subsec:LG}

\begin{figure}[t!]
\centerline{\includegraphics[angle=-90,width=0.72\columnwidth]{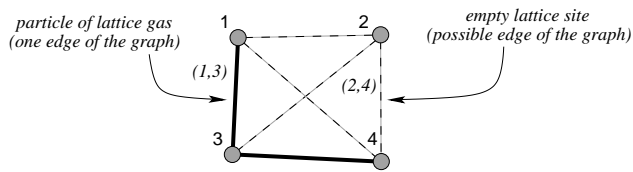}}
\caption{
Mapping a graph onto a lattice gas. 
One edge of the graph corresponds to one particle.
There are $N(N-1)/2$ possible locations for an edge in a graph:
these locations correspond to the sites of the lattice.
}
\label{fig:LG}
\end{figure}

A simple,
natural mapping of a graph with $N$ vertices onto a lattice 
gas with $N(N-1)/2$ lattice sites\footnote{This 
lattice is called the edge-dual graph of the complete 
(fully connected)
graph of $N$ vertices \cite{bollobas-book}. One lattice site
corresponds to an edge in the complete graph.
} is shown in Fig. \ref{fig:LG}.
One particle of this lattice gas corresponds to one edge of 
the original graph,
and can be at any of the $N(N-1)/2$ lattice sites. 
Two lattice sites are neighbors, 
if the corresponding two edge locations
(not necessarily occupied by edges)
of the original graph have one end point in common.
Note that this lattice strongly differs from the lattices 
generally used for lattice gases.
Taking an arbitrary edge of the graph, 
there are $2(N-2)$ other possible edges sharing an end
point with this edge: in the lattice gas, therefore, each site
has $2(N-2)$ first neighbors. All the other 
$N(N-1)/2-2(N-2)-1$ sites are second neighbors.

The quadratic single-vertex energy
is analogous to the usual definition of the energy
for a lattice gas with nearest neighbor attraction,
\bea
E=-\sum_{(\alpha,\beta)} n_\alpha n_\beta =\
-\sum_{i=1}^{N} \f{k_i(k_i-1)}{2} \, ,
\label{eq:E-LG}
\eea
\noindent 
a standard choice to describe the nucleation of vapors.
Here $n_\alpha$ is the occupation number of 
lattice site $\alpha$,
which is $0$ or $1$, depending on whether the corresponding 
edge exists in the original graph.
The summation in the
first sum goes over all pairs of neighboring particles in the lattice
gas, which corresponds to all pairs of edges 
sharing an end point
in the original graph.

This analogy can be extended to an
Ising model with a Kawasaki-type dynamics, 
where spins have $s_\alpha=2n_\alpha-1=\pm 1$ 
values and from Eq. (\ref{eq:E-LG}) the energy of the system is
\bea
E=-\sum_{(\alpha,\beta)}\f{s_\alpha+1}{2}\,\f{s_\beta+1}{2} \, ,
\label{eq:E-Ising-1}
\eea
\noindent
which can be written as
\bea
E=-\f{1}{4}\sum_{(\alpha,\beta)} s_\alpha s_\beta \
- \f{1}{2}\sum_{\alpha=1}^{N(N-1)/2} s_\alpha \
- \f{N(N-1)(N-2)}{8} \, . 
\label{eq:E-Ising}
\eea

Mapping the equilibrium graph ensemble
with the $-\sum_i k_i^2$ energy onto a lattice gas 
shows that the only difference between 
this equilibrium graph ensemble and a lattice gas 
with the nearest neighbor attraction 
$E=-\sum_{(\alpha,\beta)} n_\alpha n_\beta$
on, e.g., a cubic lattice is the {\it underlying lattice}.

\subsection{Ensembles of degenerate graphs}
\label{subsec:degenerate}

Degenerate graphs occur in almost all kinds of real-world networks, 
e.g., in food webs (cannibalism),
biochemical interaction networks 
(autocatalytic or multiple reactions), 
technological networks 
(multiple connections between subunits), 
collaboration networks 
(repeated co-authorships),
and also in field theoretic expansions of particle interactions 
in the form of Feynman graphs \cite{bessis}.

Ensembles of degenerate graphs
\cite{dm-book,dms-principles,burda-ensemble}
can be introduced similarly to the case of simple graphs.
The {\it microcanonical ensemble} 
on the set of all labeled degenerate graphs
can be defined by assigning the same weight 
to each graph with $N$ vertices and $M$ edges. 
The number of these elements can be given as follows.
There are $N(N+1)/2$ possible locations for an edge in a degenerate
graph: one can pick two different vertices to be connected by an edge
in $N(N-1)/2$ different ways, and the number of locations for
self-connections is $N$. 
Each of the $M$ (distinguishable) edges can be placed into any of
these possible $N(N+1)/2$ locations yielding
\bea
P^{\mathrm{\,MC}}=\left( \f{N(N+1)}{2}\right) ^{-M} \, 
\eea
for the microcanonical probability distribution.

It is straightforward to define a 
{\it microcanonical ensemble on a subset}
of labeled degenerate graphs. 
Since the degree distribution 
is a characteristic property of most real-world graphs, 
it can be used to select a subset: 
labeled degenerate graphs with $N$ vertices and a 
{\it fixed degree distribution}, $p_k$,
meaning that for each value of $k$ 
there are exactly $N(k)=Np_k$ 
vertices in the graph with that degree.
A given degree distribution is realized by
many adjacency matrices, 
and each adjacency matrix is further realized by
many labeled degenerate graphs 
(because the edges are distinguishable).

Since each graph has the same weight in the microcanonical ensemble,
the probability of a given adjacency matrix, $\mathbf{A}$, 
is proportional to 
the number of different graphs, ${\mathcal N}(\mathbf{A})$,
that realize this particular adjacency matrix\footnote{The
$M$ edges of the graph can be permuted in $M!$ ways.
There are $A_{jk}!$ equivalent permutations of the edges between 
vertices $j$ and $k$, but they all represent the very same graph.
Similarly, there are $(A_{ii}/2)!$ 
such equivalent permutations of the unit loops at vertex $i$.}:
\bea
P^{\mathrm{\,MC}}(\mathbf{A})\propto\
\mathcal{N}(\mathbf{A}) = \
M!\,\,
\prod_{i=1}^{N}\f{1}{(A_{ii}/2)!} \,\
\prod_{j<k=1}^{N}\f{1}{A_{jk}!}  \,.
\eea

Dorogovtsev et. al \cite{dms-principles}
have constructed canonical ensembles
of degenerate graphs by 
equilibrium processes that keep
the degree distribution and the number of edges fixed.
At each step of such a process 
one end of a randomly chosen edge is moved to a 
new vertex, $i$, selected with a weight $w(k_i)$.
Similarly, the removal of edges 
(with a rate $\lambda N$) 
together with the insertion of new edges 
between vertices $i$ and $j$ 
(with a rate proportional to $w(k_i)w(k_j)$)
lead to grand canonical ensembles.

\section{Main features of equilibrium graphs: local and global properties}
\label{sec:features}

In this section, the characteristic features of equilibrium graph
ensembles will be discussed. 
We will start with local properties
and will proceed towards properties taking 
into account larger groups of vertices.

\subsection{Local correlations}
\label{subsec:local-corr}

Most networks obtained from experimental data
contain significant correlations.
Therefore, it is a natural requirement that the models describing 
them should also contain correlated quantities. 
The frequent occurrence of connections between vertices of similar
properties such as, e.g., similar degrees, 
has been termed {\it assortativity}, 
and the higher probability of
connections between vertices with different degrees was termed 
{\it disassortativity}.
In social and biological networks, 
both assortativity and disassortativity
have been observed \cite{maslov-specificity,newman-mixing}.

One possible way of constructing
a random graph with a given degree-degree correlation,
$p(k,k')$, is the following \cite{boguna-hidden}.
First, the degree distribution, $p_k$, 
of such a graph has to be determined from
\bea
\sum_{k'} p(k,k') = \f{kp_k}{\kav} \, ,
\eea
\noindent
where $\kav=\sum_{k} kp_k$ is a condition for self-consistence.
Next, one needs to assign a random number, $q_i$,
to each vertex $i$ from the degree distribution, $p_k$.
Finally, one should go through 
each pair of vertices, $i$ and $j$, in the graph
and put a link between them with probability 
\bea
\f{\kav}{N} \f{p(q_i,q_j)}{p_{q_i}\,p_{q_j}} \, .
\label{eq:pij}
\eea
\noindent
A short technical comment here is that 
not all $p(k,k')$ functions can ensure that the
degree-degree correlation of the networks constructed with this
algorithm converges to $p(k,k')$ in the $N\to\infty$ limit.
The necessary condition is that $p(k,k')$ 
should decay slower than
$\exp(-\sqrt{k}-\sqrt{k'})$ \cite{dorog-clustering}.

An alternative approach could be to generate 
a canonical ensemble with a cost function (energy)
to suppress deviations from the prescribed $p(k,k')$ 
(see Sec. \ref{subsubsec:energy-C}).

\subsection{Global characteristics}
\label{subsec:global}

\subsubsection{Component sizes.}

One of the often studied global properties
of networks has been the size of the largest component.
Whenever the number of vertices in this component, $s_\mathrm{max}$, 
is in the order of the total number of vertices,
it is called the ``giant component''.
In the classical random graph, 
the giant component appears at the
{\it critical edge density}, $\kav =1$.
(see Ref. \cite{bollobas-book} and Sec. \ref{subsec:energy}).
Below this density the largest component contains 
${\mathcal O}(\log N)$
vertices and above this density
it will start to grow linearly. 

In a random graph with a 
{\it fixed degree distribution}, $p_k$,
the condition for the giant component to exist is 
\cite{newman-arbitPk,molloy-reed}
\bea
\sum_{k=3}^{N} k(k-2)p_k > p_1 \, .
\eea
At the transition point, the component size distribution of 
a random graph with any fixed degree distribution
is known to decay as a power law 
with the exponent $-3/2$ \cite{newman-arbitPk}.
Near the transition, the component size distribution follows
a power law with an exponential cutoff.
This is in analogy with percolation phenomena,
where the component sizes also have 
a power law distribution at the critical point.
An analytic treatment of connected components in
random graphs with fixed degree sequences is available in Ref. 
\cite{chung-components}.

\subsubsection{Spectral properties.}

Work related to the spectral properties of random structures
was launched by Wigner's semicircle law \cite{wigner}.
His result enabled the modeling of complex quantum mechanical systems
lying far beyond the reach of exact methods and later it was
found to have numerous applications in statistical and solid
state physics as well \cite{mehta,vulpiani}.
As one particular extension of Wigner's work,
F\"uredi and Koml\'os \cite{furedi-komlos}
proved that the spectral density of a classical random graph also
converges to a semicircle.
It is important to note that in the classical random graph the number
of edges is $pN^2/2$ with $p$=const., 
i.e., it grows quadratically with the number of vertices.
The general form of the semicircle law valid 
for the classical random graph \cite{petz}
states that the spectral density of 
${\mathbf A}/\sqrt{pN}$, apart from the largest eigenvalue,
will converge to
\bea
\rho(\lambda) = \
\cases {
(2\pi)^{-1} \sqrt{4-\lambda^2},& if \
$|\lambda | < 2$;\cr 0,&otherwise.}
\label{eq:semicircle}
\eea
The largest eigenvalue is detached from the 
rest of the spectrum, and scales as $pN=\kav$,
while the second largest one is about
$2 \sqrt{pN}= 2 \sqrt{\kav}$ \cite{juhasz,cvet-rowl}.
Note that reducing the density of edges may destroy the
semicircular distribution. 
In the case of a {\it sparse}\footnote{Sparse 
  graphs are a common version of graphs 
  expressing the fact that there is a cost associated to 
  each connection, therefore, the average degree
  is limited even when $N\to\infty$.
} 
classical random graph (i.e., with a fixed average degree)
$\rho(\lambda)$ converges to a distribution 
rich in singularities \cite{ifarkas-spectra,bauer-spectra}.

\unitlength10mm 
\begin{figure}[t]
\centerline{\includegraphics[angle=-90,width=0.9\columnwidth]{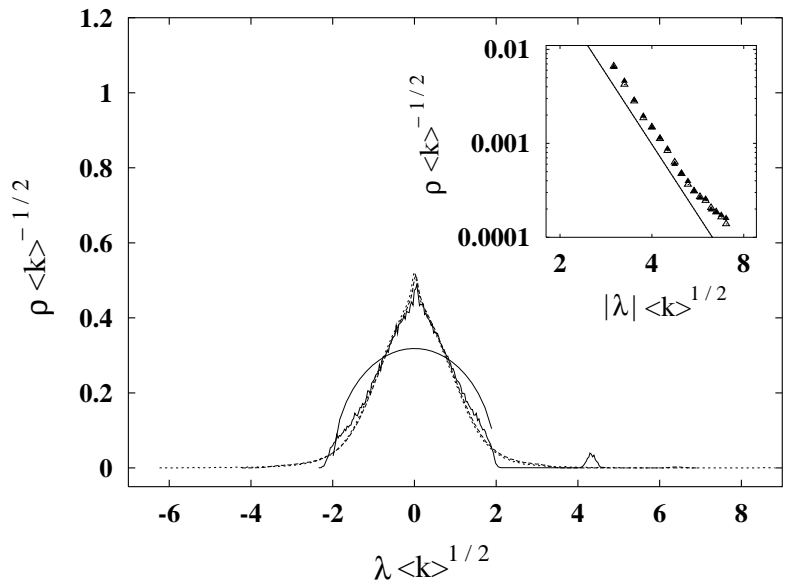}}
\caption[]{
Average spectral densities of scale-free graphs. 
(The average degree is $\langle k\rangle = 10$.)
{\bf Main panel:}
Graphs with $N=100$ (---), $N=1000$ \hbox{(-- --)},
and $N=7000$ \hbox{(- - -)} vertices
and a degree distribution decaying as 
$p_k\propto k^{-\gamma}$ ($\gamma=3$).
A continuous line shows the semi-circular distribution 
for comparison. 
The central part of the scale-free graph's 
spectral density is spiked in contrast to the flat 
top of the semi-circle.
Also, the scale-free graph's spectrum decays as a power law,
while the semicircular distribution 
decays exponentially at its edges \cite{bronx}.
{\bf Inset:} 
The upper and lower tails of $\rho(\lambda)$ 
(open and full triangles) 
for scale-free graphs with $N=40,000$ vertices.
Note that both axes are logarithmic and 
$\rho(\lambda)$ has a power law tail with the same decay rate at both
ends of the spectrum.
The line with the slope $-5$ (i.e., the exponent $\gamma=5$)
in this figure is a guide to the eye, and at the same time
a numerical prediction also
that was later confirmed by analytic results
\cite{papadimitriou,chung-spectra,dorog-spectra}.
Figure from Ref. \cite{ifarkas-spectra}.
\label{fig:scalefree-spectra}
}
\end{figure}

The next class of networks to be analyzed is 
graph ensembles with a fixed power law degree distribution.
For both real-world networks 
and graph models having a power law degree distribution,
the overall shape of the spectral density differs from the semicircle
and the largest eigenvalues follow a power law distribution
\cite{ifarkas-spectra,faloutsos,goh-spectra}
(see Fig. \ref{fig:scalefree-spectra}).
Chung et. al \cite{chung-eigvals} have found that 
a fixed power law degree distribution with the exponent $\gamma$ can be
analytically connected to a power law 
tail of the spectral density with the exponent $\alpha$:
\bea
\alpha = 2\gamma-1,\qquad\mathrm{if}\qquad\gamma > 2.5 \, .
\label{eq:chung-spec}
\eea
The findings of related numerical and analytical studies 
\cite{ifarkas-spectra,papadimitriou,dorog-spectra}
are in agreement with this result.
Evidently, the large eigenvalues are caused by the large degrees in
the graph. 
More precisely, it can be shown that 
the largest eigenvalues
can be approximated by the square roots of the largest degrees
\cite{chung-spectra}.
We mention here that the
spectral properties of a graph are closely related to random walks on
the graph 
and to the electrical resistance of the graph as a network of resistors.
For a concise review we refer the reader to Ref. \cite{lovasz-review}.

\section{Topological phase transitions in equilibrium network ensembles}
\label{sec:tpt}

As already mentioned, a widely studied phase transition
in an equilibrium network ensemble
is the occurrence of the giant component in the classical random graph
model
as a function of the density of edges. 
For $\kav <1$,
there is a similar transition in the ensemble with the 
$E=-s_{\mathrm{max}}$ energy
as a function of the temperature (see later).
An appropriate order parameter for such transitions 
is the normalized {\it size of the largest component},
$\Phi_s=s_{\mathrm{max}}/M$.
In transitions where a condensation 
of edges onto one vertex (or a small number of vertices) 
occurs, the normalized {\it largest degree}, $\Phi_k=k_{\mathbf{max}}/M$,
is the most appropriate order parameter.
In general, such transitions where
some global statistical property 
of the topology changes
(measured by an order parameter),
will be referred to as {\it topological phase transitions}.

\subsubsection{Ensembles with single-vertex energies.}
\label{subsubsec:single-vertex-energy-ensembles}

\begin{figure}[t!]
\centerline{\includegraphics[angle=-90,width=0.76\columnwidth]{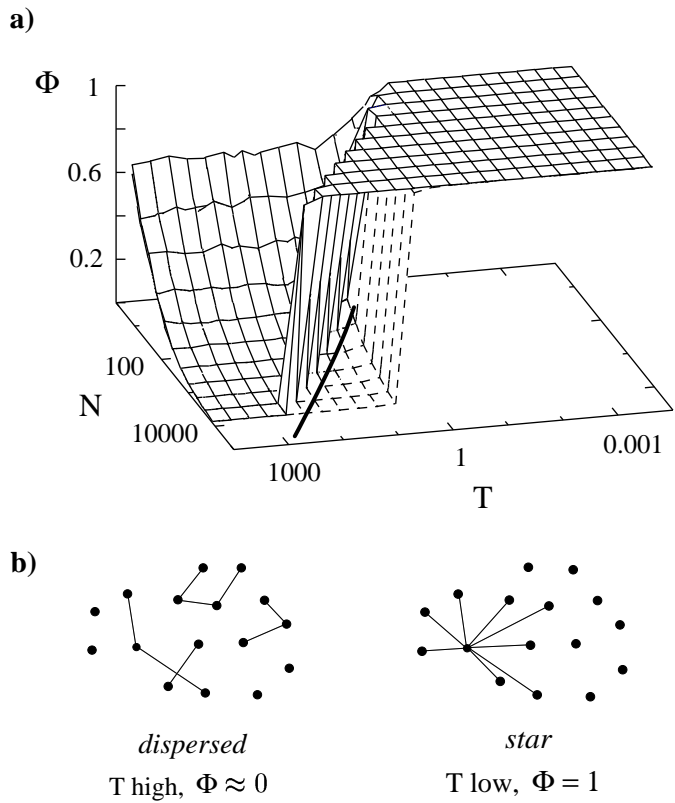}}
\caption{
Topological phase transition in the graph ensemble defined by the 
$E=-\sum_i k_i^2/2$ graph energy.
(a) The order parameter $\Phi=\Phi_k=k_{\rm max}/M$ as a function of
the temperature and the system size ($\langle k\rangle=0.5$).
The simulations were started either from a star 
(corresponding to $T=0$, solid line)
or a classical random graph 
(corresponding to $T=\infty$, dashed line). 
Each data point represents a single run, 
and averaging was carried out between the simulation times of 
$t=100N$ and $200N$ Monte-Carlo steps. 
The thick solid line shows the analytically calculated
spinodal $T_1=M/\log(N)$.
This panel is from Ref. \cite{palla}.
(b) Two typical graphs from the two phases of the graph ensemble.
At low temperatures, edges are condensed onto one vertex
($\Phi=1$), and the total energy of the system is {\it non-extensive}: 
it scales as $N^2$.
At high temperatures, 
one has a dispersed classical random graph with $\Phi\approx 0$,
and the total energy of the system scales as $N$.
\label{fig:ksqr-energy}
}
\end{figure}

For several decreasing single-vertex energies
(e.g., $E=-\sum_i k_i^2$),
a dispersed-connected phase transition can be observed 
as the temperature is changed.
In the $T\to\infty$ limit 
the dynamics converges to a completely random rewiring
process (independent of the energy function chosen),
and the classical random graph ensemble is recovered.
At lower temperatures,
since the decreasing nature of the energy function rewards 
high degrees, 
new phases appear with vertices of macroscopic degrees.

\begin{figure}[t!] 
\centerline{\includegraphics[angle=-90,width=0.8\columnwidth]{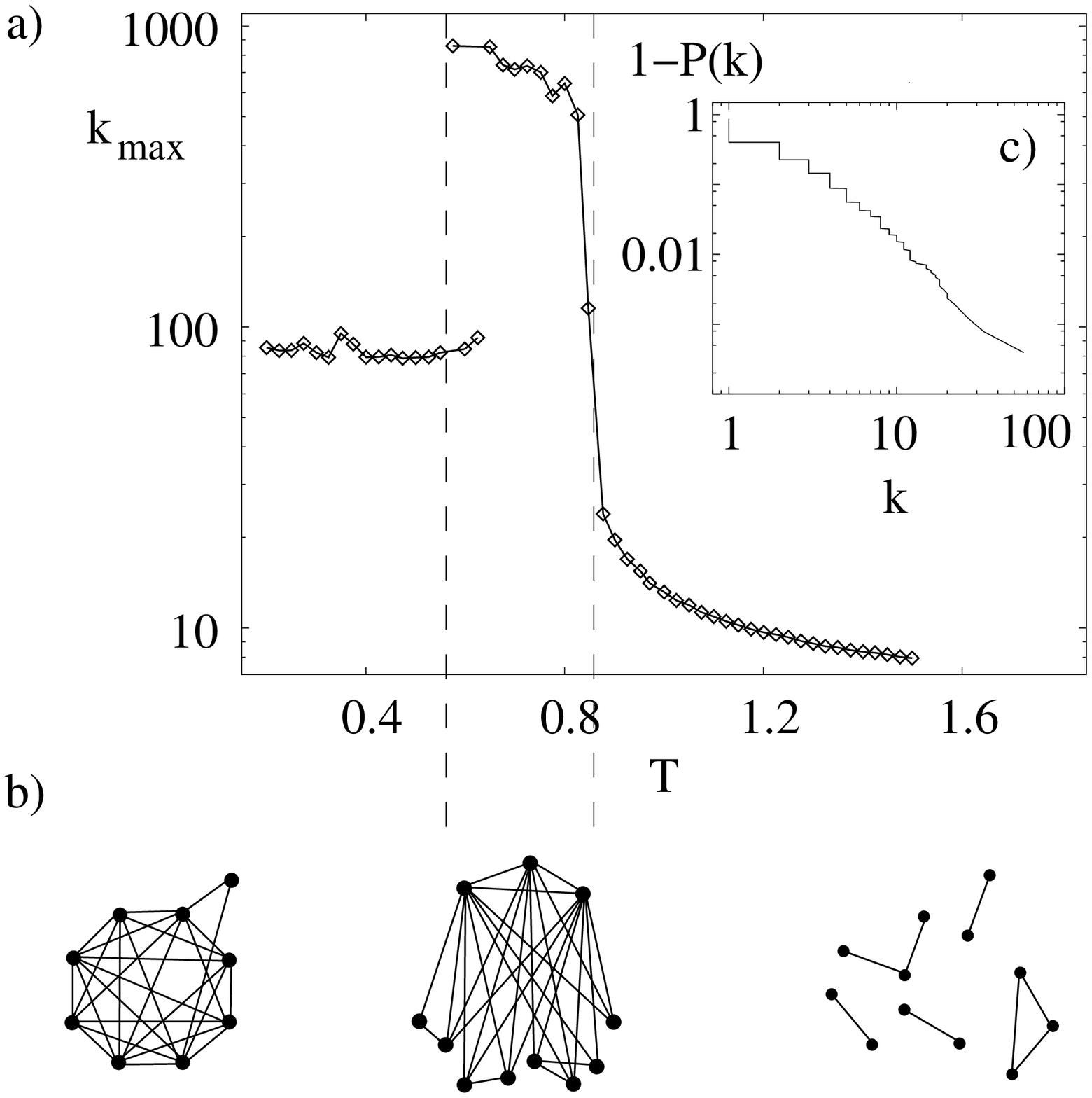}}
\caption{
Topological phases of the graph ensemble defined 
by the energy $E=-\sum_i k_i\log(k_i)$.
(a) The largest degree, $k_{\rm max}$,
for $N=10,224$ vertices and $M=2,556$ edges.
Each data point shows the value of $k_{\rm max}$ 
averaged in one simulation run between the simulation times of
$t=5,000N$ and $20,000N$ MC steps.
The data points are connected to guide the eye.
There is a sharp, continuous transition near
$T=0.85$ and a discontinuous transition (with a hysteresis) 
around $T=0.5-0.6$.
(b) The three different plateaus in (a)
correspond to distinct topological phases:
$k_{\rm max}={\mathcal O}(1)$ to the classical random graph,
$k_{\rm max}={\mathcal O}(M)$ to the star phase
(a small number of stars sharing most of their neighbors)
and 
$k_{\rm max}={\mathcal O}(\sqrt{M})$ to the fully connected subgraph.
(c) At $T=0.84$ and $t=600N$,
one minus the cumulative degree distribution, 
i.e., $1-P(k)=\int_0^k dk' p_{k'}$,
follows a power law, thus, the degree distribution decays as a power
law also.
Figure from Ref. \cite{palla}.
\label{fig:klnk}  
}
\end{figure}

Analytic calculations for 
the $E=-\sum_i k_i^2$ energy with $\kav<1$
show that between the classical random graph phase (at high
temperatures) and a phase with a star (at low temperatures)
there exists a finite intermediate temperature range 
where both phases are stable or metastable.
This predicted discontinuous transition \cite{derenyi}, 
has been confirmed by Monte-Carlo simulations 
(see Fig. \ref{fig:ksqr-energy}).
In the graph ensemble defined by the
$E=-\sum_i k_i \log (k_i)$ energy, 
two phase transitions can be observed.
Both analytical and numerical results \cite{palla} 
support that when the temperature is lowered,
the classical random graph first collapses
onto a small number of stars accompanied by a jump in the order
parameter, $\Phi_k$ (see Fig. \ref{fig:klnk}a).
In fact, in the $N\to\infty$ limit
this is a second order transition with an
infinitely large critical exponent at $T=T_{\mathrm{c}}=1$.
Further lowering the temperature will lead to another transition:
a compactification where
all edges collapse onto the minimum possible number of
vertices (see Fig. \ref{fig:klnk}b). 
This transition is discontinuous with a hysteresis. 

Note that for both single-vertex energy functions 
discussed here the total energy of the system in the
different topological phases scales differently with $N$,
which is connected to the singular changes in the average degree
(see caption of Fig. \ref{fig:ksqr-energy}).

\subsubsection{Transient ensembles vs. growing networks.}
\label{subsubsec:transient}

Non-equilibrium processes, such as, growth,
can produce a high variety of network ensembles.
Some of these ensembles can also be constructed 
with the help of an equilibrium dynamics
as a {\it transient ensemble}, 
i.e., as an intermediary and temporary
ensemble between an initial set of graphs and
the final, equilibrium ensemble.
In the case of the $E=-\sum_i k_i \log (k_i)$ energy,
during the process of relaxation from the classical random graph phase
to the star-like phase near the critical temperature, $T_{\mathrm{c}}=1$
(see Fig. \ref{fig:klnk}b for two typical
graphs illustrating these two phases),
the degree distribution of the graphs in the transient ensemble
decays continuously, as a power law
(see Fig. \ref{fig:klnk}c).
The qualitative description of this phenomenon is the following.
During the transition there is a ``pool'' of edges attached to vertices
of small degrees, and a small number of vertices with higher degrees
serve as centers of condensation.
The change of energy associated with moving edges within the ``pool''
is negligible, whereas the nucleation centers are accumulating
edges at a rate proportional to their degrees\footnote{If 
  an edge from
  the ``pool'' is moved to a vertex with a large degree ($k$), 
  then the
  energy of the system changes by approximately 
  $\Delta E=\partial E/\partial k = -\log k -1$.
  In the equilibrium dynamics the rate of this step
  will be $\e^{-\Delta E / T}\propto k$ 
  at the critical temperature,
  $T=T_{\mathrm{c}}=1$. 
}.
This mechanism, produced by an equilibrium dynamics,
is analogous to the 
preferential attachment rule of growing (non-equilibrium)
models of scale-free networks \cite{sf,sf2}, 
which also lead to power law degree distributions.

\subsubsection{Ensembles with neighboring vertex energies.}

Baiesi and Manna \cite{manna-hamiltonian}
have analyzed the canonical ensemble of connected graphs
defined by the energy shown in Eq. (\ref{eq:E-manna}). 
This energy favors degree dissasoratativity, i.e., a negative
degree-degree correlation. 
As a function of temperature three phases have been identified 
in this ensemble:
the classical random graph at $T\to\infty$,
scale-free graphs at intermediate temperatures,
and a phase with a small number of stars 
at low temperatures.

\subsubsection{Ensembles with component energies.}

\begin{figure}[t!]
\centerline{\includegraphics[angle=-90,width=0.9\columnwidth]{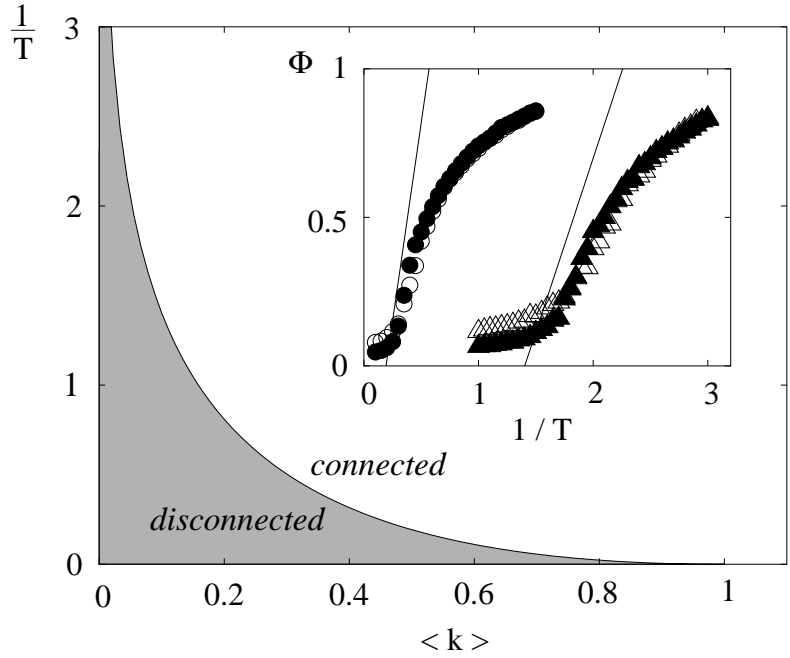}}
\caption{
Analytical phase diagram and Monte-Carlo simulation results 
for the graph ensemble defined by the $E=-s_{\rm max}$ energy.
{\bf Main panel:}
The white and shaded areas correspond to the ordered phase
(containing a giant component) and the disordered phase, respectively
as given by Eq. (\ref{eq:smax-line}).
{\bf Inset:} 
The order parameter $\Phi=\Phi_s=s_{\rm max}/M$
obtained from Monte-Carlo simulations
as a function of the inverse
temperature for $\langle k\rangle=0.1$ (triangles) and 
$\langle k\rangle=0.5$ (circles).
Each data point shows averages taken 
for $10$ runs between the simulation times of 
$t=100N$ and $500N$ Monte-Carlo steps. 
The open and closed
symbols represent $N=500$ and $1,000$ vertices, respectively.
The critical exponent, in agreement with the
analytical approximations (solid lines, see Ref. \cite{derenyi}), 
was found to be $1$. 
Figure from Ref. \cite{palla}.
\label{fig:smax}
}
\end{figure}

Similarly to the single-vertex case, a decreasing, component-size
dependent energy can also lead to phase transitions.
The simplest case which we analyze in this paragraph, 
is $E=-s_{\rm max}$.
At low densities ($\kav<1$)
one can observe a classical random graph at $T\to\infty$,
whereas at low temperatures a giant component is present.
It can be shown that the 
dividing line between the two topological phases 
is \cite{palla}
\bea
T_{\rm c}(\kav)=\f{1}{\kav-1-\log(\kav)} \, ,
\label{eq:smax-line}
\eea
which is also supported by numerical results 
(see Fig. \ref{fig:smax}).
In the vicinity of the critical temperature the
order parameter can be approximated as
\bea
\Phi_s^*(T)=2 \, \f{T^{-1}-T^{-1}_{\rm c}(\kav)}{\kav^2-3\kav+2} \, ,
\eea
indicating that the phase transition is
{\it continuous}
(see Fig. \ref{fig:smax} for details).

\subsubsection{Further ensembles.}

For the ensembles of degenerate graphs
introduced by Dorogovtsev et. al \cite{dms-principles} 
(see Sec. \ref{subsec:degenerate}),
with $w(k)\propto k +1 -\gamma$, 
a critical line, $\kav=k_{\mathrm{c}}(\gamma)$, was found. 
Below this line the degree distribution has an exponential
cutoff and above that a {\it condensate} occurs where
a finite fraction of all edges is attached to an inifitely small
fraction of nodes.

\begin{figure}[t!]
\centerline{\includegraphics[angle=0,width=0.6\columnwidth]{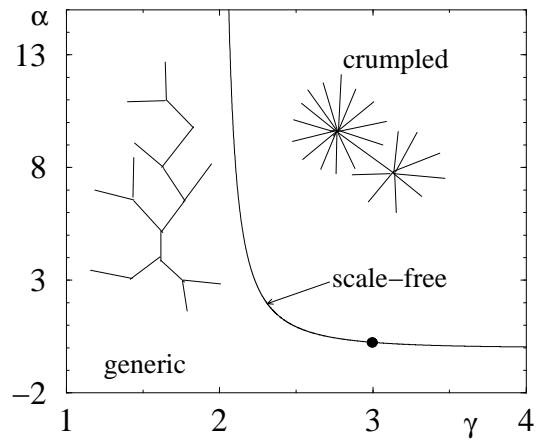}}
\caption{
Phase diagram of the two-parameter ensemble of scale-free tree graphs
presented in Ref. \cite{burda-ensemble}.
Two phases were identified: {\it generic} and {\it crumpled}.
The points of the dividing line 
to the right from the dot are scale-free graphs 
and belong to the generic phase.
Figure from Ref. \cite{burda-ensemble}.
\label{fig:burda}
}
\end{figure}

In an ensemble of connected tree graphs with 
a fixed, power law degree sequence,
Burda et. al \cite{burda-ensemble} have reported a 
phase transition as a function of two parameters:
$\gamma$, the exponent of the degree distribution,
and $\alpha$, 
related to the probability of subgraphs
(see Fig. \ref{fig:burda}).
The analytic form of the dividing line between the identified
{\it generic} and {\it crumpled} phases was computed, 
and numerical simulations were carried out 
using a Monte-Carlo sampling technique.

\section{Summary}
\label{sec:sum}

Graph models with energies provide a natural way to define
microcanonical, canonical and grand canonical ensembles.
These ensembles are often generated by equilibrium restructuring 
processes obeying detailed balance and ergodicity.
Also, to describe a wider range of network models,
it is useful to extend the definition and consider
ensembles without energy as well.
We have reviewed the main features of currently studied
equilibrium graph ensembles, 
with a focus on degree-degree correlations, 
component sizes and spectral properties.
We have also discussed continuous and discontinuous topological phase
transitions in equilibrium graph ensembles.
A solid basis of the equilibrium statistical 
mechanics of networks, as presented in this article,
can facilitate the application of statistical physics tools
in the field of networks, and can help to 
expand the analyses towards problems of high current
interest, such as optimization and reverse engineering.

\section{Acknowledgements}
\label{sec:ack}

This work has been in part supported by 
the Hungarian Scientific Research Fund
under grant No: OTKA 034995. 
I. F. acknowledges a scholarship from the
Communication Networks Laboratory at ELTE.


\end{document}